\documentclass[sn-aps]{sn-jnl}

\usepackage{graphicx}%
\usepackage{multirow}%
\usepackage{amsmath,amssymb,amsfonts}%
\usepackage{amsthm}%
\usepackage{mathtools}%
\usepackage{mathrsfs}%
\usepackage[title]{appendix}%
\usepackage{xcolor}%
\usepackage{textcomp}%
\usepackage{manyfoot}%
\usepackage{booktabs}%
\usepackage{algorithm}%
\usepackage{algorithmicx}%
\usepackage{algpseudocode}%
\usepackage{listings}
\usepackage{orcidlink}
\usepackage[export]{adjustbox}
\usepackage[normalem]{ulem}

\begin{document}

\title[Article Title]{
The Nordic-walking mechanism and its explanation of deconfined pseudocriticality from Wess-Zumino-Witten theory
}

\author[1]{\fnm{Bilal} \sur{Hawashin}\orcidlink{0009-0005-6335-0483}}\email{bilal.hawashin@rub.de}

\author[2]{\fnm{Astrid} \sur{Eichhorn}\orcidlink{0000-0003-4458-1495}}\email{eichhorn@cp3.sdu.dk}

\author[3]{\fnm{Lukas} \sur{Janssen}\orcidlink{0000-0003-4919-796X}}\email{lukas.janssen@tu-dresden.de}

\author[1]{\fnm{Michael M.} \sur{Scherer}\orcidlink{0000-0003-0766-9949}}\email{scherer@tp3.rub.de}

\author*[2]{\fnm{Shouryya} \sur{Ray}\orcidlink{0000-0003-4754-0955}}\email{sray@cp3.sdu.dk}

\affil[1]{\orgdiv{Institut f\"ur Theoretische Physik III}, \orgname{Ruhr-Universit\"at Bochum}, \orgaddress{\postcode{D-44801} \city{Bochum}, \country{Germany}}}

\affil[2]{\orgdiv{CP3-Origins}, \orgname{University  of  Southern  Denmark}, \orgaddress{\street{Campusvej  55}, \city{5230 Odense M}, \country{Denmark}}}

\affil[3]{\orgdiv{Institut f\"ur Theoretische Physik and W\"urzburg-Dresden Cluster of Excellence ct.qmat}, \orgname{TU Dresden}, \orgaddress{\postcode{01062} \city{Dresden}, \country{Germany}}}


\abstract{The understanding of phenomena falling outside the Ginzburg-Landau paradigm of phase transitions represents a key challenge in condensed matter physics. 
A famous class of examples is constituted by the putative deconfined quantum critical points between two symmetry-broken phases in layered quantum magnets, such as pressurised SrCu\textsubscript{2}(BO\textsubscript{3})\textsubscript{2}. 
Experiments find a weak first-order transition, which simulations of relevant microscopic models can reproduce.
The origin of this behaviour has been a matter of considerable debate for several years.
In this work, we demonstrate that the nature of the deconfined quantum critical point can be best understood in terms of a novel dynamical mechanism, termed \emph{Nordic~walking}. 
Nordic walking denotes a renormalisation group flow arising from a beta function that is flat over a range of couplings.
This gives rise to a logarithmic flow that is faster than the well-known \emph{walking} behaviour, associated with the annihilation and complexification of fixed points, but still significantly slower than the generic \emph{running} of couplings. 
The Nordic-walking mechanism can thus explain weak first-order transitions, but may also play a role in high-energy physics, where it could solve hierarchy problems.

We analyse the Wess-Zumino-Witten field theory pertinent to deconfined quantum critical points with a topological term in 2+1 dimensions.
To this end, we construct an advanced functional renormalisation group approach based on higher-order regulators.
We thereby calculate the beta function directly in 2+1 dimensions and provide evidence for Nordic walking.}

\maketitle

\section{Introduction}

The Ginzburg-Landau paradigm of phase transitions is remarkable in its seemingly universal applicability. 
Consequently, finding and thoroughly understanding examples that violate this paradigm is invaluable to discovering new mechanisms and their applications to quantum condensed matter physics and beyond.
A proposed mechanism beyond Ginzburg-Landau theory that has been a matter of intense debate over the last two decades~\cite{Senthil:2004deconfined, Shao:2016quantum, Wang:2017deconfined, Cui:2023proximate, Senthil:2023deconfined} has fractionalised degrees of freedom emerge at a quantum phase transition between two long-range-ordered states that break different symmetries. 
The fractionalisation mechanism circumvents the Ginzburg-Landau restrictions and, in principle, would allow a continuous transition without fine tuning.
Recently, the mechanism has been argued to be at work in the layered quantum magnet SrCu$_2$(BO$_3$)$_2$ under hydrostatic pressure~\cite{Guo:2023deconfined, Cui:2023proximate}.
However, both in experiments, as well as in simulations of microscopic models, the transition appears to be discontinuous, albeit only very weakly so~\cite{Nahum:2015deconfined, Zhou:2023so5, song2023deconfined}.
It has been argued that the correct theory to describe this behaviour is a non-linear (2+1)-dimensional sigma model together with a level-$k$ Wess-Zumino-Witten (WZW) term~\cite{PhysRevLett.95.036402,PhysRevB.74.064405}. %
As $k$ increases and reaches $k_\text{W}$, the infrared stable Renormalisation Group (RG) fixed point of the non-linear sigma model annihilates with a second, unstable fixed point. %
Subsequently both become complex~\cite{Nahum:2015deconfined, Wang:2017deconfined,Nahum:2020note,Ma:2020theory}.
Complex fixed points with very small imaginary part result in a very slow RG flow in the real space of couplings, causing an exponentially large correlation length and a suppression of discontinuities at the transition~\cite{Kaplan:2009conformality, Gukov:2017RG}.
This scenario has been termed \emph{walking}~\cite{Gorbenko:2018walkingI, Gorbenko:2018WalkingII}, in order to distinguish the behaviour from the usual \emph{running} of couplings in the RG.

The walking scenario can only explain the phenomenology observed in the experiments and the simulations, if the physical value of $k$ is above, but not too far from the critical $k_\text{W}$.
However, a robust calculation of $k_\text{W}$ and a quantitative characterisation of the RG flow for $k < k_\text{W}$ has been hampered by the perturbative non-renormalisability of the WZW field theory and by the topological character of the theory, which prevents a standard dimensional continuation based on an epsilon expansion~\cite{Ma:2020theory, Nahum:2020note}.
As a consequence, the level-$k$ WZW theory has remained theoretically well understood only in the weak-coupling region~\cite{Senthil:2023deconfined}.

In this work, we demonstrate that the functional renormalisation group (FRG) framework allows us to overcome these challenges.
We thus for the first time robustly test the walking scenario for the deconfined quantum phase transition.
We find that while annihilation and complexification of fixed points indeed occurs, the resulting value for $k_\text{W}$ is too far from the physical value $k=1$ and cannot explain the drifting exponents and the weak nature of the transition observed in simulations.
We instead discover a new mechanism that explains a large correlation length through a dynamical change of a scaling dimension to near marginality.
This results in a logarithmically slow drifting of critical exponents with, however, a larger coefficient as compared to the walking scenario. %
Consequently, we dub the mechanism \emph{Nordic walking}. %

\section{Walking versus Nordic walking}

Consider a statistical or quantum field theory with a coupling $g$. %
Upon renormalisation, it depends on the infrared cutoff scale $\kappa$ described by the beta function~$\beta_g$, i.e., the coupling flows according to $\kappa \,\partial_\kappa g(\kappa) = \beta_g(g(\kappa))$. %
The beta function depends on the theory under consideration; here, we assume that it can be written in the form 
\begin{align}
    \beta_g(g;C) = g - g^2 + C g^n\,.
    \label{eq:betag_example}
\end{align}
In the above, $n > 2$ is an integer, $C$ is a positive external parameter, and $g =g(\kappa)$ has canonical mass dimension $[g]=-1$, as indicated by the linear contribution in~$g$.
In Sec.~\ref{sec:results}, we specifically provide the beta function of WZW theory, which has a similar form to Eq.~\eqref{eq:betag_example}.

\begin{figure}[!tb]
\centering
\includegraphics[scale=0.8]{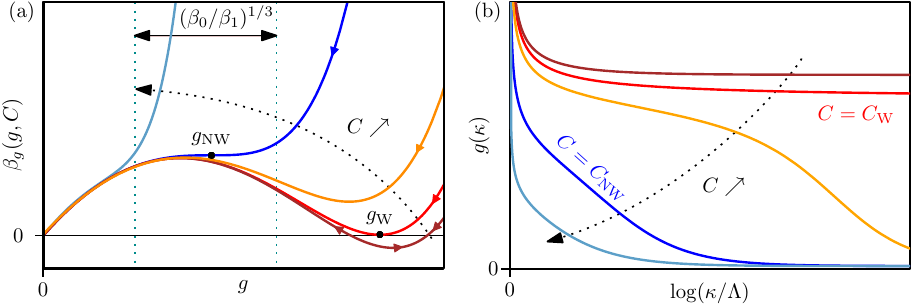}
\caption{\label{fig:nordicwalking}
{\bf Schematic RG flow, illustrating walking and Nordic walking mechanisms.}
(a)~Graphs of $\beta_g(g;C)$ for different values of $C$, from Eq.~\eqref{eq:betag_example} with $n = 6$. 
(b)~Corresponding flow of the coupling starting from a large value of $g$ in the UV.
For small $C$ below $C_{\text{W}}$ (brown curve), there are two real roots $g_{*,0}$ and $g_{*,1}$, of which the latter is a stable fixed point.
When $C$ approaches $C_{\text{W}}$ (red curve), these two merge to a double root. For $C$ above, but close to, $C_{\text{W}}$ (orange curve), there are no real roots left, but the beta function is small at its local minimum $g \approx g_\text{W}$, leading to an arbitrarily large correlation length.
When $C$ approaches $C_\text{NW}$ (blue curve), the local maximum and minimum of the beta function annihilate at $g = g_\text{NW}$.
For $C$ above, but close to, $C_{\text{NW}}$, the beta function is strictly monotonous, but still approximately flat around its inflection point, leading to a large correlation length, see text. 
For $C \gg C_\mathrm{NW}$ (turquoise curve), the beta function is steep, leading to a large scale dependence of the coupling.
The ratio $(\beta_0/\beta_1)^{1/3}$ determines the width of the Nordic walking regime, with $\beta_0$ and $\beta_1$ the coefficients in the expansion of the beta function around $g_\text{NW}$, see Eq.~\eqref{eq:betanwvinc}.
}
\end{figure}

RG fixed points are real roots of $\beta_g$. %
They define scale-invariant field theories, such as those realised at a second-order phase transition, with concomitant critical behaviour at long wavelengths.
For $C$ small enough, $\beta_g$ in Eq.~\eqref{eq:betag_example} has two positive real roots, located at $g_{*,0}$ and $g_{*,1}$, see Fig.~\ref{fig:nordicwalking}.
We are interested in weaker, approximate notions of scale invariance, that may mimic true scale invariance in experiments.

\subsection{Walking}

For increasing $C$, the two fixed points approach each other and coalesce at $g=g_\text{W}$ at a critical value of $C = C_{\text{W}}$. 
For $C > C_{\text{W}}$, there are no real fixed points, because the fixed points form a complex conjugate pair $g_{*,\pm}$. 
As a consequence, if $C$ is just slightly above $C_\text{W}$, the beta function at $g= g_\text{W} \approx \operatorname{Re}g_{*,\pm}$ 
is close to zero, see Fig.~\ref{fig:nordicwalking}. 
Thus, the flow near $g(\kappa) \approx \operatorname{Re} g_{*,\pm}$ is slow; 
the system is approximately scale-invariant.
This slow running is referred to as \emph{walking} \cite{Kaplan:2009conformality, Gukov:2017RG, Gorbenko:2018walkingI, Gorbenko:2018WalkingII}.
If the real fixed points for $C$ below $C_{\text{W}}$ describe a second-order phase transition, the walking behaviour for $C$ above, but close to $C_{\text{W}}$, corresponds to a weak-first-order transition with an exponentially large correlation length and an exponentially small jump in the order parameter.

The fixed-point annihilation mechanism and its concomitant slow RG flow is currently considered by many~\cite{Nahum:2015deconfined, Wang:2017deconfined, Ma:2020theory, Nahum:2020note, Senthil:2023deconfined} as the best explanation of the quasi-universal weakly-first-order behaviour observed in experiments~\cite{Cui:2023proximate, Guo:2023deconfined} and simulations~\cite{Nahum:2015deconfined, song2023deconfined} of the deconfined quantum phase transition.
The mechanism itself, however, is quite general and has been discussed in a variety of systems, including quantum electrodynamics in $2+1$ spacetime dimensions~\cite{braunPhase2014, giombiConformal2016, janssenSpontaneous2016, herbutChiral2016, Gukov:2017RG}, the Abelian Higgs model~\cite{halperin1974first, ihrig2019Abelian}, many-flavour quantum chromodynamics~\cite{giesChiral2006, Kaplan:2009conformality, Gukov:2017RG, bondUV2018, bondConformal2021}, technicolour and compositeness~\cite{Foadi:2007ue, Cacciapaglia:2020kgq}, higher-dimensional scalar field theories~\cite{fei2014critical, fei2015three, eichhorn2016critical}, tensor models~\cite{herbutCritical2016, graceyTensor2018}, Potts models~\cite{Gorbenko:2018WalkingII, maShadow2019}, long-range-interacting fermion models~\cite{herbutTopological2014, janssenPhase2017}, theories with multiple order parameters~\cite{gehringFixed2015, herbutMulticriticality2022, ladovrechisGross2023, uetrecht2023absence}, and quantum impurity models~\cite{weberSU2023}.

\subsection{Nordic walking}

The salient feature of the beta function in the Nordic-walking scenario is that $\beta_g \approx \beta_0 = \rm const$.
When the beta function is constant, the coupling runs logarithmically, because 
\begin{equation}
 \kappa \partial_{\kappa}g(\kappa)\sim
 \beta_0 \implies g(\kappa) \sim g(\kappa_\text{i})+ \beta_0
 \ln \left(\frac{\kappa}{\kappa_\text{i}}\right),
 \label{eq:genericNordicwalking}
\end{equation}
where $\kappa_\text{i}$ is a reference scale.

A concrete example of how a beta function can give rise to this type of behaviour is illustrated by our toy-model beta function, Eq.~\eqref{eq:betag_example}.
Here, Nordic walking is realised when $C$ is significantly above $C_\text{W}$, but close to another critical value $C_\text{NW}$, at which the beta function features a saddle point at $g_\text{NW}$, see Fig.~\ref{fig:nordicwalking}. 
For values of $C$ near $C_\text{NW}$, the beta function is approximately constant over an extended range of couplings around $g_\text{NW}$. 
Because the original coupling has mass dimension $-1$, its canonical scale-dependence is a linear power-law scaling. 
Instead, a logarithmic scale dependence would be associated to dimensionless, i.e., marginal couplings. 
The fact that the beta function becomes constant can thus be understood as a dynamical marginalisation of the coupling.
This mechanism of dynamical marginalisation systematically slows down the generic RG running of the coupling, but it is still faster than in the case of walking.
Hence, we name it \emph{Nordic walking}.

Nordic walking can also lead to a large correlation length, as we will argue below, and thus it can potentially explain the results of numerical simulations on deconfined criticality. The extent of the Nordic-walking regime and the size of the correlation length depend on the numerical values of the coefficients in the beta function.
The smaller the constant $\beta_0$ in Eq.~\eqref{eq:genericNordicwalking}, the slower the logarithmic scale dependence.
However, even for a large value of $\beta_0$, the scale dependence remains logarithmic, even if the canonical scale-dependence of the coupling is much stronger, e.g., for a canonically relevant or irrelevant coupling. 
As an important difference to critical scaling or walking, having a Nordic-walking regime does therefore not rely on the absolute value of the beta function being sufficiently small. 
The size of the constant $\beta_0$, does, however, influence the length of a Nordic-walking regime: 
the larger $\beta_0$, the shorter the Nordic-walking regime. 
Thus, $\beta_0$ also enters the correlation length. 
Further, the correlation length depends on the next-to-leading term in the beta function, which controls the width of the flat region in the beta function, as we will explain now.

The mechanism that leads to the flattening of the toy-model beta function can be understood to arise from another type of collision, in contrast to the fixed-point collision underlying the conventional walking scenario:
For $C$ approaching $C_{\text{NW}}$ from below, the local extrema of the beta function come closer and eventually collide to merge into the saddle point at $g_{\text{NW}}$.
At $C_\text{NW}$ and in the vicinity of the saddle point, the beta function can generically be brought into the form
\begin{align}\label{eq:betanwvinc}
    \beta_g(g,C_\text{NW}) &= \beta_0 + \beta_1 (g - g_{\text{NW}})^3+\ldots \,,
\end{align}
where $\beta_0>0$. 
For convenience, we also assume $\beta_1 > 0$.
The region, in which the beta function is approximately constant within a, say, $10\%$ relative range around $\beta_0$ is determined by $\lvert g-g_\text{NW} \rvert \lesssim \frac{1}{2} (\beta_0/\beta_1)^{1/3}$.
Integrating the flow of $g$ from an initial scale $\kappa_\text{i}$, at which it enters the flat region, to a final scale $\kappa_\text{f}$, at which it exits the region, with $g(\kappa_{\text{i},\text{f}}) \simeq g_\mathrm{NW} \pm \frac{1}{2} (\beta_0/\beta_1)^{1/3}$, we find for the ratio of scales
\begin{equation}\label{eq:fracNW}
    \frac{\kappa_\text{i}}{\kappa_\text{f}} \simeq \exp \left((\beta_0^2 \beta_1)^{-1/3} \right).
\end{equation}
For $C$ near, but not precisely at, $C_\text{NW}$, a similar form for $\kappa_\text{i}/\kappa_\text{f}$ can be derived. In this case, an additional factor $\exp[a(C-C_\mathrm{NW})]$ appears, with nonuniversal constant~$a$.
As in the conventional walking scenario, one can interpret the ratio of scales $\kappa_\text{i}/\kappa_\text{f}$ as the dimensionless correlation length $\xi$ in units of the inverse ultraviolet cutoff $a_0 \sim 1/\Lambda$, which is naturally given on the lattice.
Equation~\eqref{eq:fracNW} now implies that there is an interplay between the value of the beta function at the saddle point, determined by $\beta_0$, and the width of the flat region, determined by $(\beta_0/\beta_1)^{1/3}$. 
For any given $\beta_0$, we obtain an exponentially enhanced correlation length, if the width of the flat region $(\beta_0/\beta_1)^{1/3}$ is large compared to $\beta_0^{-1/3}$.

At this point, some comments are in order:
First, we note that the only condition for Nordic walking behaviour is a wide region in which the beta function is approximately flat.
As this is equivalent to the requirement of (nearly) vanishing derivatives of the beta function, Eqs.~\eqref{eq:betanwvinc} and \eqref{eq:fracNW} describe the generic behaviour arising in any system that exhibits Nordic walking.

Second, while the toy beta function of Eq.~\eqref{eq:betag_example}, as well as the WZW theory discussed below, display both Nordic walking and standard walking for different parameter values $C$, other systems do not need to walk before they Nordic walk. The only condition for Nordic walking is a sufficiently wide region in which the beta function is flat, which does not require a preceding fixed-point annihilation to take place.

Third, Nordic walking may be an important mechanism in condensed-matter systems as well as high-energy physics.
In condensed matter, it is of particular interest for canonically irrelevant couplings, which vanish as a power-law under the RG flow to the IR. 
In order for such a coupling to play a role in the low-energy physics, it has to be enhanced. Nordic walking, because it slows a power-law decay to a logarithmic decay, can achieve such an enhancement. 
The system has to be sufficiently non-perturbative such that higher-order terms in the beta function can generate the required flatness.
In high-energy physics, Nordic walking can reduce or even solve hierarchy problems. 
These are generically associated to canonically relevant couplings and arise because such a dimensionful coupling is much smaller than the typical mass scale of the theory. 
In the standard setting, in order to accommodate this hierarchy, one fine-tunes the initial value of the coupling at a UV scale, such that the initial value compensates the large contribution from radiative corrections and generates a small value of the coupling in the IR. 
If instead the coupling exhibits Nordic walking, the contribution from radiative corrections is much smaller, because it builds up logarithmically instead of as a power-law.
Accordingly, no fine-tuning of the initial value of the coupling at the UV scale is necessary in order to generate a hierarchy in the theory.

In summary, we have (i)~scale invariance, if the beta function vanishes, (ii)~walking, i.e., arbitrarily slow scale dependence, if the first derivative of the beta function (nearly) vanishes and and the beta function is close to zero (iii)~Nordic walking, i.e., a logarithmic scale dependence, if the first and the second derivative of the beta function (nearly) vanish so that the beta function is approximately constant, even if it is not close to zero.

\section{Results}\label{sec:results}

\subsection{Model}

We consider WZW theory in $2+1$ spacetime dimensions with target manifold $\operatorname{SO}(5)/\operatorname{SO}(4) \cong S^4$, described by the action $S = S_{\text{SO}(5)} + 2\pi ik\Gamma_{\text{WZW}}$ with
\begin{align}
    S_{\text{SO}(5)} &= \frac{1}{4\pi \bar{g}} \int_{\mathbb{R}^3} d^3 x \, (\partial_\mu \Phi(x))^2, \label{eq:origSO5}\\
    \Gamma_{\text{WZW}} &= \frac{1}{64 \pi^2}\!\! \int_0^1\!\!\! du\! \int_{\mathbb{R}^3}\!\! d^{3}x \, \epsilon^{M_1\ldots M_{4}}\epsilon^{a_0\ldots a_{4}} \Phi_{a_0}(x,u) \partial_{M_1}\Phi_{a_1}(x,u) \ldots  \partial_{M_{4}}\Phi_{a_{4}}(x,u). \label{eq:origWZWterm}
\end{align}
The field $\Phi = (\Phi_a) = (\phi_1,\phi_2)^\top$ has five components and is constrained by $|\Phi|^2 = 1$. 
In the application to deconfined quantum criticality, $\phi_1$ corresponds to the N\'eel order parameter $\phi_1 = (n_x, n_y , n_z)$ and $\phi_2$ to the valence bond solid (VBS) order parameter $\phi_2 = (V_1,V_2)$~\cite{PhysRevLett.95.036402,PhysRevB.74.064405}, see Ref.~\cite{Senthil:2023deconfined} for a recent review.

In the above, $S_{\text{SO}(5)}$ describes the dynamics of four massless modes in the four-dimensional target manifold $S^4$ in terms of a nonlinear sigma model.
Physically, the modes can be understood near the deconfined quantum critical point as follows: 
Two modes become the Goldstone modes in the N\'eel phase, corresponding to the breaking of $\operatorname{SO}(3)$ $(\cong \operatorname{SU}(2)/\mathbb{Z}_2)$ spin symmetry. 
Another one becomes the Goldstone mode in the VBS phase, corresponding to the breaking of XY lattice rotational symmetry.
The remaining fourth mode is, roughly speaking, the one that corresponds to rotating the N\'eel order parameter into the VBS order parameter and vice versa. 
It is gapped in the presence of finite anisotropy~$|\phi_1|^2 - |\phi_2|^2$.
This anisotropy tunes the N\'eel-VBS transition. Right at the transition, the renormalised anisotropy is zero, i.e., it drops out of the action.%
\footnote{The anisotropy does, however, control leading critical exponents. For instance, its scaling dimension yields the correlation-length exponent.}
The parameter $\bar g$ corresponds to the coupling in the nonlinear sigma model, with mass dimension $[\bar g] = -1$.
The WZW term $\Gamma_\text{WZW}$ is constructed by embedding the three-dimensional spacetime in a four-dimensional bulk manifold $\mathbb{R}^3 \times [0,1]$, with $\partial_{M} = (\partial_\mu,\partial_u)$ denoting the four-dimensional gradient.
The field variable is promoted to $\Phi(x,u)$ with boundary conditions $\Phi(x,0) = (1,0,0,0,0)$ and $\Phi(x,1) = \Phi(x)$.
The functional integral arising from the above action is independent of the precise nature of the embedding, if and only if the coefficient $k$, i.e., the WZW level, is quantised as $k \in \mathbb{Z}$. 
The physical case of the N\'eel-VBS transition on the context of deconfined quantum criticality corresponds to $k=1$~\cite{Senthil:2023deconfined}.

\subsection{Flow equation}

For a generic value of the coupling $\bar{g}$, the WZW theory is not scale invariant. 
This is quantified by the theory's beta function, which tracks the flow of the coupling as a function of RG scale $\kappa$, $\beta_{g} = \kappa \partial_\kappa g$. 
We have here introduced the dimensionless avatar $g$ of the coupling $\bar{g}$ appearing in Eq.~\eqref{eq:origSO5} as $g(\kappa) = \kappa \bar{g}(\kappa)$. 
The fact that $g$ is dimensionful behooves us to choose a Wilson--Kadanoff `mode decimation' approach to RG. 
In such approaches, the RG scale $\kappa$ is implemented as an infrared cut-off. 
At the level of the path integral measure, this may be expressed schematically as the modification
\begin{align}
    \int (d\Phi) e^{-S} \to \int (d\Phi) e^{-S - \Delta S_\kappa}, \qquad \Delta S_\kappa = \int d^3 x \left(r(-\partial^2/\kappa^2)\,\partial_\mu \Phi_a\right)^2. 
    \label{eq:shapefunctiondef}
\end{align}
The function $r(\cdot)$ in the above equation is called shape function; 
the freedom of choosing different $r$'s corresponds roughly to the choice of different schemes in perturbative RG. 
The effective action computed with the $\kappa$-dependent measure is the flowing effective action, denoted $\Gamma_\kappa$ \cite{Wetterich:1992yh}, see \cite{Dupuis:2020fhh} for a recent review.
There are multiple advantages of implementing the RG scale $\kappa$ as an infrared cut-off. Already on a conceptual level, this ensures that beta functions of the theory fully encode the variation of the effective action under global scale transformations, viz., through the Ward--Takahashi identity for global scale transformations \cite{Morris:2018zgy}; 
for a general implementation of RG, this is not guaranteed in the presence of dimensionful couplings.
On the computational level, the flow $\kappa \partial_\kappa \Gamma_\kappa$ can be expressed self-consistently in terms of functional derivatives of $\Gamma_\kappa$. 
Consequently, one can set up self-consistent approximations that do not rely on the smallness of the interaction strength $g$. 
This allows us to construct an ansatz for $\Gamma_\kappa$ using well-known principles of effective field theory.

For our computation, we employ the minimal truncation $\Gamma_\kappa = S|_{\bar{g} \to g(\kappa)/\kappa}$. 
We implement the constraint $|\Phi|^2 = 1$ exactly by a suitable non-linear choice of field variables \cite{Cardy:1996xt, Ma:2020theory},
\begin{align}
    \Phi_\alpha &= \tau_\alpha, \quad \alpha \in \{ 1,2,3 \}, \label{eq:parm1}\\
    \Phi_4 &= \sqrt{1 - \vec\tau^2} \cos \vartheta, \\
    \Phi_5 &= \sqrt{1 - \vec\tau^2} \sin \vartheta,
    \label{eq:parm-1}
\end{align}
We keep terms up to quartic order, $O(\{\vartheta,\vec{\tau}\}^4)$. 
This leads to non-linear terms in $\Delta S_\kappa$ in Eq.~\eqref{eq:shapefunctiondef} and induces additional contributions to the standard Ellwanger--Morris--Wetterich equation, see Ref.~\cite{Pawlowski:2005xe} for the general regulator theory. 
We finally arrive at the beta function (see Methods section for details and a diagrammatic representation)
\begin{align}
    \beta_g = g + \frac{- a_1(r) g^2 + a_2(r)k^2 g^6}{1 - b_1(r) g + b_2(r) k^2 g^5}.
    \label{eq:betafct}
\end{align}
In the above, the coefficients $a_i$ and $b_i$ depend on the shape function $r$. 
For sufficiently well-behaved regulators, all the coefficients $a_i$ and $b_i$ are positive (see Methods section for details).
The terms proportional to $a_1$ and $b_1$ describe the usual flow in the $(2+1)$-dimensional O(5) nonlinear sigma model~\cite{Codello:2008qq,Flore:2012ma,Efremov:2021fub}.
The nontrivial terms proportional to $a_2$ and $b_2$ capture the physically important feedback of the WZW level $k$ (which is quantised and does not flow) on the flowing coupling $g(\kappa)$. 
The non-trivial denominator arises from the non-perturbative nature of the flow equation, but is in fact not important at the qualitative level; our most important results can thus be obtained from a series expansion of the beta function.

We use optimisation techniques to determine a shape function $r$ that is expected to be least sensitive to truncation artefacts (see Methods section for details). The coefficients $a_i$ and $b_i$ of such optimised shape function read
\begin{align}
 a_1^\text{opt} \approx 0.90,\quad
     a_2^\text{opt} \approx 0.88,\quad
b_1^\text{opt} \approx 0.22, \quad
    b_2^\text{opt} \approx 0.19.
\end{align}
%

\subsection{Critical WZW levels for walking and Nordic walking}

To discover whether walking is an explanation of deconfined criticality, we first work with a simplified version\footnote{A different simplification, in which the denominator is neglected completely, leads to results which are in qualitative agreement with our analysis.} of the beta function Eq.~\eqref{eq:betafct}, where we Taylor-expand the beta function to order $g^6$:
\begin{equation}
 \beta_g= g - 0.90 g^2 - 0.20 g^3 - 0.044 g^4 - 0.0096 g^5 + 
 g^6 (-0.0021 + 0.88 k^2)\,.
\end{equation}
The fixed-point structure and resulting scale dependence depends on the WZW level $k$, which we treat as an external parameter.

At $k=0$, there is only one fixed point at finite positive coupling, located at $g_{*,0} = 0.89$. It has one infrared relevant direction within the SO(5) theory space.
This fixed point can be identified with the critical point in the usual nonlinear sigma model (without topological term), whose existence has been studied extensively in $D = 2+\epsilon$ dimensions using perturbative methods \cite{Brezin:1975sq,BrezinPRD76,Brezin:1976qa,Bardeen:1976zh,Bernreuther:1986js,Wegner:1989ss,Hikami:1980yg}, as well as for different fixed $D$ using functional methods \cite{Codello:2008qq,Flore:2012ma,Efremov:2021fub}.

For small finite $k$, another real interacting fixed point at $g_{*,1}>0$ emerges from infinite coupling. 
It can be interpreted as the three-dimensional avatar of the WZW fixed point in two spacetime dimensions, if one imagines the existence of a continuous family of WZW fixed points in $D$ spacetime dimensions, with target manifold $\operatorname{SO}(D+2)/\operatorname{SO}(D+1)$ and $D$ as a continuous parameter.%
\footnote{We stress, however, that a controlled way to smoothly interpolate between the $D=2$ and $D=3$ WZW theories is not known.}
In $D=2$, the theory is exactly soluble for integer $k$ \cite{Polyakov:1983tt,Witten:1983ar,Knizhnik:1984nr}, and the fixed point is located at $g_{*,1}(D=2) \propto 1/k$.

The critical value $k_{\text{W}}$ at which the fixed points $g_{\ast, 0}$ and $g_{\ast, 1}$ collide is
$k_\text{W} \approx 0.467$. 
For $k$ not much above $k_\text{W}$, the coupling $g$ walks. Because the physical WZW level $k=1$ is significantly above our estimate for $k_\text{W}$, our calculation does \emph{not} support the conjecture that walking is the mechanism behind deconfined pseudocriticality.

Instead, we discover that there is a $k_\text{NW}$ at which the beta function becomes approximately constant and the coupling instead exhibits Nordic walking. 
For Nordic walking, the slope of the beta function at the inflection point changes its sign. 
This is determined by
\begin{align}
    \frac{\partial \beta_g}{\partial g} (g_{\text{NW}};k_{\text{NW}}) = 0, 
    \qquad 
    \frac{\partial^2 \beta_g}{\partial g^2} (g_{\text{NW}};k_{\text{NW}}) = 0,
\end{align}
where $g_{\text{NW}}$ corresponds to the location of the inflection point. 
We obtain a critical value $k_\text{NW} \approx 0.91$ that is rather close to the physical value $k=1$ of the WZW level. 
This supports Nordic walking as the mechanism behind deconfined pseudocriticality.

\begin{figure}[tb]
    \centering
    \mbox{\includegraphics[scale=0.75]{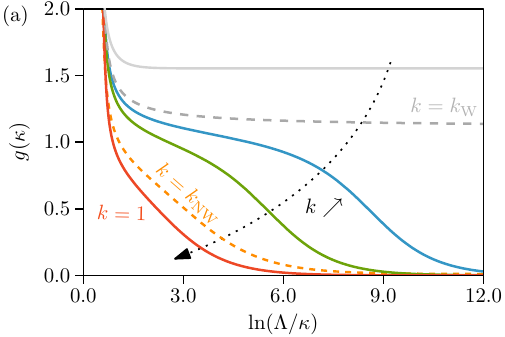}\hfill%
    \includegraphics[scale=0.75]{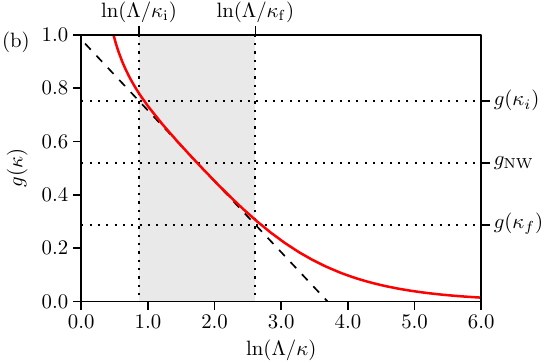}}
    \caption{
    \label{fig:flow_g_all_ks}
    {\bf Flow of coupling $g$ as function of RG scale $\kappa$ for different fixed values of the WZW level $k$.}
    (a)~For $k$ below $k_\text{W}$, there is true scaling behaviour in the infrared $\ln(\Lambda/\kappa) \to \infty$ (light grey curve).
    For $k$ just above $k_{\text{W}}$, there are several orders of magnitude in $\kappa$ where $g(\kappa) \approx \operatorname{Re} g_{*,\pm}$ (blue curve). For increasing $k$, this walking regime shrinks. For the physical value $k = 1$ (red curve), no walking regime is left. However, there is now instead a Nordic walking regime during which $g(\kappa)$ is approximately linear in $\ln (\Lambda/\kappa)$.
    (b)~Zoom into the Nordic regime for $k=1$. The Nordic walking regime, highlighted in grey, is centred around $\beta_g$'s inflection point, $g = g_{\text{NW}}$, and has a width of approximately $2$ orders of magnitude in $\kappa$, $\kappa_\text{i}/\kappa_\text{f} \approx \mathrm e^2$. 
    For instance, if Nordic walking begins to manifest at linear system size $L_\text{i} \sim 64$, it will carry on until about $L_\text{f} \sim \mathrm{e}^2 L_\text{i} \sim 512$.
    }
\end{figure}

Going beyond the approximate form of our beta function and instead using its full form in Eq.~\eqref{eq:betafct}, we find\footnote{There is a small numerical uncertainty in the result of about $10\,\%$ arising from the statistical error in the numerical integration, but given the systematic uncertainties, we abstain from quoting it all the time.}
\begin{align}
k_\text{W} &\approx 0.4, \label{eq:kWopt} &
k_\text{NW}& \approx 0.8.
\end{align}

We show the resulting scale dependence of $g$ for different values of $k$ in Fig.~\ref{fig:flow_g_all_ks}(a), where it is immediately apparent that $k=1$ is far away from the walking regime characterised by a very slow change of $g$ with $\kappa$. Instead, the physical case $k=1$ exhibits Nordic walking, i.e., the dependence of $g$ on $\ln \left(\Lambda/\kappa \right)$ is linear.
We zoom in on the Nordic walking regime in Fig.~\ref{fig:flow_g_all_ks}(b), where the flow for $k=1$ exhibits logarithmic scaling for nearly two decades. 
In this regime, we have
\begin{align}
    g(\kappa) \approx g(\kappa_\text{i})  - v_g \ln (\kappa_\text{i} / \kappa)
\end{align}
where $g(\kappa_\text{i})$ corresponds to the onset of Nordic walking and we have introduced the drift velocity $v_g > 0$. Numerically, we find $v_g\approx 0.3$. 

We expect that scaling dimensions drift with velocities that are specific to a given operator at hand, but only differ from $v_g$ by factors of $\mathcal{O}(1)$ (and potentially signs). Indeed, lattice results calculate the correlation length exponent $\nu$ at two different lattice spacings and find $\nu_\text{Num} = 0.62$ at $L = 64$  and  $\nu \approx 0.48$ for $L = 256$ \cite{Nahum:2015deconfined}. This works out to a drift velocity of roughly $v_\nu\approx 0.14/\ln 4 \approx 0.23$, which is indeed of a similar magnitude as our estimate for the drift velocity of the coupling in the Nordic-walking regime. We caution that our comparison of $v_g$ and $v_\nu$ is based on a plausibility argument, but we do not actually extract the scaling dimension of the two-point function from our analysis. This could in the future be attempted by using a flow equation for composite operators \cite{Pawlowski:2005xe,Igarashi:2009tj,Pagani:2016dof}.

In the Nordic-walking regime, we evaluate the scaling dimension associated to the coupling $g$, which is given by
\begin{align}
    \Delta_\mathrm{S}(\kappa) = 3 + \left.\frac{\partial\beta_g}{\partial g}\right|_{g=g(\kappa)},
\end{align}
which we evaluate as a function of $\kappa$, because we do not have a fixed-point regime. When $\Delta_{\rm S}$ dips below the spacetime dimension of 3, this irrelevant operator becomes relevant. This indeed happens on trajectories for which $k < k_\text{NW}$, but is avoided for the Nordic-walking regime when $k > k_\text{NW}$, including the physical case $k=1$: 
Because $g$ is dynamically marginalised, and $\operatorname{S}$ is the operator that couples to $g$, $\Delta_\text{S}$ does get close to~3 in that regime. 
However, $\Delta_\text{S}$ never drops below~3, i.e., the least-irrelevant scalar perturbation remains irrelevant throughout.

\begin{figure}
    \centering
    \includegraphics[scale=0.75]{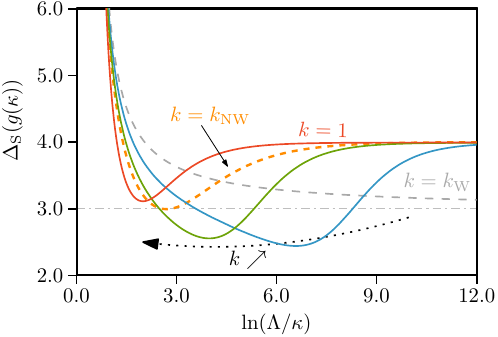}
    \caption{
    {\bf Effective scaling dimension of least-irrelevant scalar.}
    For $k > k_{\text{NW}}$, $\Delta_{\mathrm{S}}(g(\kappa))$ remains greater than the spacetime dimension, i.e., it is \emph{not} dangerously irrelevant in the strict sense. For $k < k_{\text{NW}}$, on the other hand, it becomes relevant during the course of the flow.}
    \label{fig:scaldimofg} 
\end{figure}

\section{Discussion}

\begin{figure}
    \centering
    \includegraphics[width=0.5\textwidth]{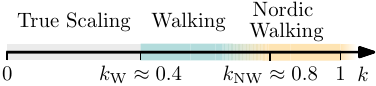}
    \caption{
    {\bf Scaling behaviour of the WZW model for different values of the level $k$.} 
    For $k < k_\text{W}$, the system features a stable fixed point at positive values of the coupling $g$, corresponding to true scaling behaviour with a divergent correlation length.
    When $k$ approaches $k_\text{W}$, the stable fixed point collides with another fixed point, and both move into the complex plane for $k> k_\text{W}$. 
    This leads to walking behaviour as long as $k$ remains close to $k_\text{W}$, characterised by a very large correlation length originating from a small value of the beta function.
    The deconfined pseudocritical point corresponds to the WZW level $k=1$, which is significantly away from $k_\text{W}$ and falls into the Nordic walking regime, characterised by a large correlation length originating from a flat (but not necessarily small) beta function.}
    \label{fig:critksschem}
\end{figure}

We have presented the first non-perturbative renormalisation group calculation of $(2+1)$D WZW theory. While we have focused on the target manifold $\operatorname{SO}(5)/\operatorname{SO}(4) \cong S^4$, the 4-sphere, our new scheme works for general classes of target manifolds. 
As a function of the WZW level $k$, our beta function demonstrates the annihilation of an unstable fixed point with the stable interacting fixed point. 
The former can be understood as the finite-$k$ avatar of the fixed point discovered within the non-linear sigma model, the latter is the 3D version of the 2D WZW CFT.
Our best guess for the critical value $k_{\text{W}}$ of the WZW level where the fixed-point annihilation occurs is roughly $k_{\text{W}} \approx 0.4$, cf.~Fig.~\ref{fig:critksschem}. 
However, despite the fact that the physical $k=1$ is rather far from $k_{\text{W}}$, we do find a sizable regime in which the flow is logarithmically slow, leading to drifting effective exponents.
This drifting of exponents arises from a new mechanism, which we have termed \emph{Nordic walking}, as an intermediate concept between \emph{walking} and \emph{running} of couplings.
Nordic walking arises when the beta function is approximately constant over a sufficiently broad interval of couplings. This can for instance originate from an annihilation of \emph{extrema} of the beta function (as opposed to its roots) at a second critical WZW level $k_{\text{NW}} > k_{\text{W}}$, see Fig.~\ref{fig:critksschem}.
We have obtained $k_{\text{NW}} \approx 0.8$, which is significantly closer to the physical $k=1$. As a result, the flow of the coupling becomes approximately logarithmic in a range of scales.
We calculate the drift velocity $v_g$ of the coupling in the Nordic-walking regime and find $v_g \approx 0.3$. It is plausible that exponents drift with velocities that only differ from $v_g$ by factors of order 1 (and possible signs). Indeed, $v_g$ comes out close to what large-scale simulations find~\cite{Nahum:2015deconfined}:
$\nu \approx 0.62$ at $L \sim 64$ and $\nu \approx 0.48$ at $L \sim 256$, with a drift velocity of $v \approx 0.23$. Whether or not this can be interpreted as evidence for Nordic walking in the lattice simulations requires further study and a calculation of $\nu$ from our flow-equation setup.
Furthermore, the least irrelevant perturbation remains irrelevant throughout the flow for $k = 1 \gtrsim k_{\text{NW}}$, unlike for smaller values of $k$ below $k_{\text{NW}}$ closer to $k_{\text{W}}$. 

Our framework admits straightforward extensions of the truncation for the flowing effective action. 
With those, one could for the first time attempt to approach apparent convergence of the critical levels $k_\text{W}$ and $k_\text{NW}$ in a theoretical calculation directly within WZW theory at fixed spacetime dimension. 
Excitingly, the fact that this does not rely on mapping the WZW theory to some dual theory opens the possibility of exploring theories where such dualities may not be known or may not even exist, with non-Lagrangian phases of matter~\cite{Zou:2021dwv} representing an interesting example.

The mechanism of Nordic walking we have proposed herein could play a role in many other systems.
Not only does it provide an explanation of pseudocritical behaviour, it can also dynamically generate hierarchies between distinct scales in a natural way. 
For instance, if an appropriate extension of the Standard Model of particle physics results in a near-flat beta function for the mass parameter of the Higgs, then a large separation between the Planck mass and the Higgs mass can emerge naturally, without any need for fine-tuned RG trajectories. 
Similarly, any other coupling that has a canonical power-law dependence can, if it exhibits Nordic walking, circumvent naturalness expectations. 
This may, for instance, be important in the context of the Standard Model Effective Field Theory~\cite{Falkowski:2023hsg}. 
In that setting, the Standard Model is extended by higher-dimensional operators, the couplings of which are not marginal. 
Thus, a natural expectation is that these couplings can only be sizeable at LHC scales, if the scale of new physics is nearby. 
However, if some of these couplings in fact exhibit Nordic walking in or beyond the Standard Model, a sizeable coupling at LHC scales can be reconciled with a far-off new-physics scale.
Nordic walking could thus have important implications in our understanding of LHC data and in model-building for theories beyond the Standard Model.

\appendix

\section{Methods}

\subsection{Functional Renormalisation Group and Higher-Order Regulator Insertions}

We employ the non-perturbative functional renormalisation group (FRG)~\cite{Reichert:2020mja,Dupuis:2020fhh}. For our purposes, its main advantage derives from the fact that the RG scale $\kappa$ is implemented as an infrared cut-off. 
Consequently, the running effective action $\Gamma_\kappa$, called \emph{average effective action}, can be expanded in a series of local operators $\mathcal{O}_i$ that are homogeneous in fields and derivatives, which constitute an eigenbasis of the dilatation operator. In conjunction with the Ward--Takahashi identity %
$\delta_\epsilon \Gamma = \delta_\epsilon \Gamma_\kappa - \kappa \partial_\kappa \Gamma_\kappa$ %
\cite{Morris:2018zgy}, this implies that scale invariance of the full effective action $\Gamma$ is precisely equivalent to the vanishing of all dimensionless beta functions of the theory. 

Consider a general theory with $N$ real scalar fields $\phi_l(x)$, $l \in \{1, \ldots ,N\}$ and a UV cut-off $\Lambda$. 
The average effective action is defined implicitly by the path integral
\begin{equation}
    e^{-\Gamma_\kappa[\phi] - \Delta S_\kappa'[\phi,R]} = \int (d\psi)\, \exp\left[-S[\phi + \psi] - \Delta S_\kappa[\phi + \psi,R] + \int_x \frac{\delta \Gamma_\kappa(\phi)}{\delta \phi(x)}\cdot \psi(x)\right],
\end{equation}
where the regulating term $\Delta S_\kappa[\phi,R]$ is assumed to be of the form
\begin{align}
    \Delta S_\kappa\left[\phi,R\right] &= \sum_{n\geq 1}\int_{x_1 \ldots  x_n} R_\kappa^{l_1 \ldots  l_n}(x_1,\ldots ,x_n) \phi_{l_1}(x_1) \cdots \phi_{l_n}(x_n),\label{eq:genregins}
\end{align}
and $\Delta S_\kappa'[\phi,R]$ refers to $\Delta S_\kappa[\phi,R]$ without the $n=1$ term.
Note that in most applications, the regulator can be taken to be bilinear in the fields, i.e., the series is taken to terminate at $n=2$ \cite{Reichert:2020mja,Dupuis:2020fhh}. We shall, however, encounter also higher-order terms, which cannot be omitted, and therefore review the general formalism~\cite{Pawlowski:2005xe}. The flow of the effective action is given by
\begin{align}\label{genfloweq}
    \dot{\Gamma}_\kappa &= \dot{R}_\kappa^{\hphantom{\kappa}A} \phi_A \\ 
    &\hphantom{=} {}+ \sum_{n\geq 2} \dot{R}_\kappa^{\hphantom{\kappa}A_1\ldots A_n} \bigg[\left( G_\kappa\frac{\delta}{\delta \phi} + \phi\right)_{A_1}\cdots \left( G_\kappa\frac{\delta}{\delta \phi} + \phi\right)_{A_{n-1}} - \phi_{A_1} \cdots \phi_{A_{n-1}}\bigg] \phi_{A_n}
\end{align}
where, for brevity, we have introduced the deWitt index $A = (l,x)$ and the overdot denotes the scale derivative $\kappa \partial_\kappa$; $G_\kappa$ denotes the full field-dependent two-point function, 
\begin{equation}
    \left(G_\kappa^{-1}\right)^{A_1 A_2} = \frac{\delta^2 (\Gamma_\kappa + \Delta S_\kappa')}{\delta \phi_{A_1}\delta \phi_{A_2}}.
\end{equation}
The general flow equation simplifies to the well-known Wetterich--Morris--Ellwanger equation \cite{Wetterich:1992yh,Morris:1993qb,Ellwanger:1993mw}, if only the $n=2$ coefficient is non-vanishing,
\begin{equation}
    \dot{\Gamma}_\kappa = \dot{R}_\kappa^{\hphantom{\kappa}A_1 A_2} \, G_{\kappa\:A_1 A_2} = \includegraphics[scale=1,valign=c]{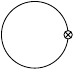}.
\end{equation}
In the diagrammatic representation, the internal line stands for $G_{\kappa\:A_1 A_2}$, and $\otimes$ for an insertion of $\dot{R}_\kappa^{\hphantom{\kappa}A_1 A_2}$. In our calculation, we shall also need the $n = 4$ term, which leads to the diagrammatic equation
\begin{align}
    \dot{\Gamma}_\kappa = \includegraphics[scale=1,valign=c]{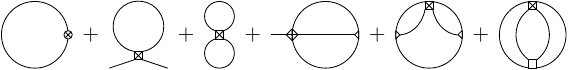},
    \label{eq:flow_master}
\end{align}
with $\boxtimes$ standing for an insertion of $\dot{R}_\kappa^{\hphantom{\kappa}A_1 A_2 A_3 A_4}$ and $\triangle$ and $\square$ denote the third and fourth derivatives of $(\Gamma + \Delta S')_\kappa$, respectively.
The general theory for what constitutes a physical regulator for $n > 2$ is quite involved and not completely studied yet \cite{Pawlowski:2005xe}. However, as we will see below, the quartic term derives from a non-linear variable transformation of a quadratic master regulator, for which the conditions are in fact well-understood by now~\cite{Gies:2006wv}.

\subsection{Truncation and Flow Equation}

The general strategy for solving the flow equation non-perturbatively is to make an ansatz -- a so-called \emph{truncation} -- for the average effective action $\Gamma_\kappa$, and solving the resulting equation self-consistently. In the present case, a subtlety arises because the SO(5) symmetry is realised non-linearly in terms of the $\vec{\tau}$ and $\vartheta$ fields. The route we shall follow is to simply write down a conventional manifestly SO(5) invariant regulator term as in Eq.~\eqref{eq:shapefunctiondef} in terms of the original $\Phi_a$ fields,
\begin{align}
    \Delta S_\kappa &= \frac{1}{4 \pi g_\kappa} \int d^3x \left( \sqrt{r(-\partial^2/\kappa^2)}\,\partial_\mu \Phi_a \right)^2,
\end{align}
and express this in terms of $\vec{\tau},\vartheta$ by inserting Eqs.~\eqref{eq:parm1}--\eqref{eq:parm-1}. Up to fourth order in the fields, this yields
\begin{align}
\Delta S_\kappa &= \frac{1}{4 \pi g_\kappa} \int_p\left[R_\kappa(p^2) \vartheta(p) \vartheta(-p)+ R_\kappa(p^2) \vec{\tau}(p) \cdot \vec{\tau}(-p)\right] \nonumber\\
& +\frac{1}{4 \pi g_\kappa} \int_{p_1, \ldots, p_4} \frac{1}{4} R_\kappa(p_{12}^2) \vec{\tau}\left(p_1\right) \cdot \vec{\tau}\left(p_2\right) \vec{\tau} \left(p_3\right) \cdot \vec{\tau}\left(p_4\right)\delta_{1234} \nonumber\\
& +\frac{1}{4 \pi g_\kappa} \int_{p_1, \ldots, p_4} \left(\frac{1}{2} R_\kappa(p_{12}^2) - R_\kappa(p_4^2) \right) \vec{\tau}\left(p_1\right) \cdot \vec{\tau}\left(p_2\right) \vartheta\left(p_3\right) \vartheta\left(p_4\right)\delta_{1234} \nonumber\\
& +\frac{1}{4 \pi g_\kappa} \int_{p_1, \ldots, p_4} \left( \frac{1}{4} R_\kappa(p_{12}^2) - \frac{1}{3} R_\kappa(p_4^2) \right) \vartheta\left(p_1\right) \vartheta\left(p_2\right) \vartheta\left(p_3\right) \vartheta\left(p_4\right)\delta_{1234} \nonumber\\
& + O(\{\vec{\tau},\vartheta\}^5),
\end{align}
with $R_\kappa(p^2) \coloneqq p^2 r(p^2/\kappa^2)$, $\delta_{1234} \coloneqq \delta^3(p_1 + \ldots + p_4)$, $p_{12} \coloneqq p_1 + p_2$, and $\int_p \coloneqq \int_{\mathbb{R}^3} d^3 p/(2\pi)^3$. As can be seen from the above, the quartic coefficients are fully determined by the quadratic coefficients.

For the average effective action $\Gamma_\kappa$, we employ a minimal truncation in that we take the effective action to be the classical action, but with the coupling $g$ promoted to a flowing coupling $g_\kappa$, i.e., 
$\Gamma_\kappa = S|_{\bar{g} \to g_\kappa/\kappa}$.
Note that a possible wavefunction renormalisation is neglected, which is sufficient to the leading order. In general, the WZW level should stay quantised under RG flow, and hence the corresponding beta function should also vanish. 
Up to fourth order in the non-linear fields, the average effective action reads 
\begin{align}
    \Gamma_\kappa &= \frac{1}{4 \pi g_\kappa} \int_x\left(\partial_\mu \vartheta\right)^2+\frac{1}{4 \pi g_\kappa} \int_x\left(\partial_\mu \tau\right)^2+\frac{1}{4 \pi g_\kappa} \int_x \tau^2\left(\partial_\mu \tau\right)^2-\frac{1}{4 \pi g_\kappa} \int_x \tau^2\left(\partial_\mu \vartheta\right)^2 \nonumber\\
    &\hphantom{=} {}+ \frac{ik}{8 \pi} \int_x \epsilon^{ \mu \nu \rho} \epsilon^{\alpha \beta \gamma} \partial_\mu \vartheta\,\tau_\alpha \partial_\nu \tau_\beta \partial_\rho \tau_\gamma,    
\end{align}
where $\int_x \coloneqq \int_{\mathbb{R}^3} d^3x$. It is instructive to collect the pertinent regularised versions of the propagators and vertices to appear in the evaluation of the master flow equation \eqref{eq:flow_master},
\begin{align}
    \includegraphics[scale=1,valign=c]{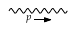} \quad &= \frac{2\pi g_\kappa}{P_\kappa(p^2)}, \\ 
    \includegraphics[scale=1,valign=c]{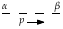} \quad &= \frac{2\pi g_\kappa}{P_\kappa(p^2)}\delta_{\alpha \beta}, \\
    \includegraphics[scale=1,valign=c]{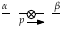} \quad & =  \frac{1}{2 \pi g_\kappa} \delta_{\alpha \beta} \left[ \dot{R}_\kappa\!\left(p^2\right) - \frac{\beta_g}{g_\kappa} R_\kappa\!\left(p^2\right)\right], \\
    \includegraphics[scale=1,valign=c]{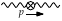} \quad & =  \frac{1}{2 \pi g_\kappa} \left[ \dot{R}_\kappa\!\left(p^2\right) - \frac{\beta_g}{g_\kappa} R_\kappa\!\left(p^2\right)\right], \\
    \includegraphics[scale=1,valign=c]{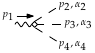} & = \frac{ik}{4 \pi} \epsilon^{\mu \nu \rho} \epsilon^{\alpha_2 \alpha_3 \alpha_4} (-ip_{4,\mu}) (p_{3,\nu}p_{4,\rho} - p_{2,\nu}p_{4,\rho}  + p_{2,\nu}p_{3,\rho} ), \\
     \includegraphics[scale=1,valign=c]{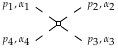} & = \frac{1}{2 \pi g_\kappa} [R_\kappa\!\left(p_{34}^2\right) \delta_{\alpha_1 \alpha_2} \delta_{\alpha_3 \alpha_4}+ R_\kappa\!\left(p_{14}^2\right) \delta_{\alpha_1 \alpha_4} \delta_{\alpha_2 \alpha_3}+ R_\kappa\!\left(p_{24}^2\right) \delta_{\alpha_1 \alpha_3} \delta_{\alpha_2 \alpha_4}], \label{eq:quarticregvertex1}\\
     \includegraphics[scale=1,valign=c]{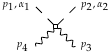} & =  \frac{1}{2 \pi g_\kappa} \delta_{\alpha_1 \alpha_2} \left[2 p_3 \cdot p_4+R_\kappa\!\left(p_{34}^2\right)-2 R_\kappa\!\left(p_3^2\right)\right] \label{eq:quarticregvertex-1}, \\ 
     \includegraphics[scale=1,valign=c]{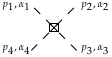} & =  \frac{1}{2 \pi g_\kappa} \left[\left(\dot{R}_\kappa\left(p_{34}^2\right) - \frac{\beta_g}{g} R_\kappa\left(p_{34}^2\right) \right) \delta_{\alpha_1 \alpha_2} \delta_{\alpha_3 \alpha_4} + (1 \leftrightarrow 3) + (2 \leftrightarrow 3)\right],\\ 
     \includegraphics[scale=1,valign=c]{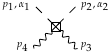} & =  \frac{1}{2 \pi g_\kappa} \delta_{\alpha_1 \alpha_2} \left[\left(\dot{R}_\kappa\left(p_{34}^2\right) - \frac{\beta_g}{g} R_\kappa\left(p_{34}^2\right) \right)-2 \left(\dot{R}_\kappa\left(p_{4}^2\right) - \frac{\beta_g}{g} R_\kappa\left(p_{4}^2\right) \right)\right] ,
\end{align}
with $P_\kappa(p^2) \coloneqq p^2 [1 + r(p^2/\kappa^2)]$ and $p_{i,\mu}$ denoting the $\mu$-th component of $p_i$.

The flow of $g_\kappa$ is fixed by the normalisation of the leading-order in momentum term of $\vartheta$'s propagator,
\begin{align}
    \frac{-\beta_g}{4\pi g_\kappa^2}\delta^3(0) = \frac{1}{12}\left.\delta_{\mu\nu}\frac{\partial^2}{\partial p_\mu \partial p_\nu}\frac{\delta^2 \dot\Gamma_\kappa[\vec{\tau},\vartheta]}{\delta \vartheta(p)\vartheta(-p)}\right|_{\vec{\tau},\vartheta,p^2 \to 0}.
\end{align}
When evaluating the master equation, we formally distinguish contributions into two categories: (i) those that do not vanish for $k \to 0$ and (ii) those that dominate for $k \to \infty$ for $g^{4+\delta} k^2$ held fixed for any $\delta > 0$, and keep the leading-order contributions to each category. Within the present truncation, this yields
\begin{align}\label{betag}
    \frac{-\beta_g}{4 \pi g_\kappa^2} = \includegraphics[scale=1,valign=c]{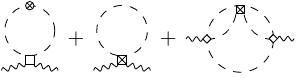},
\end{align}
where for the last diagram, we only keep the quartic WZW vertex. Importantly, note that the quartic regulator insertion is \textit{required} to capture the effect of the WZW term within our truncation.

Evaluating the diagrams, we obtain the beta function of the dimensionless coupling
\begin{equation}
    \beta_g = g + \frac{- C_{1,4}(r) g^2 + C_{1,20}(r)k^2 g^6}{1 - C_{0,4}(r) g + C_{0,20}(r) k^2 g^5}.
\end{equation}
with coefficients
\begin{align}
    C_{1,4}(r) &= 3 I_{12}(r) + 4 \pi I_{15}(r), \\
    C_{0,4}(r) &= 3 I_{02}(r) + 4 \pi I_{05}(r), \\
    C_{1,20}(r) &= \frac{9}{16 \pi^2} (2 \pi)^5 I_{15}(r), \\
    C_{0,20}(r) &= \frac{9}{16 \pi^2} (2 \pi)^5 I_{10}(r),
\end{align}
%
%
in terms of the threshold integrals
\begin{align}
\label{eq:threshold11}
    I_{11}(r) &=  \frac{4 \pi}{3} \int_{l} \bigg[ \frac{ r'\left(l^2\right)}{(1+r(l^2))^2} \left(7 l^2 r_\text{reg}'\left(l^2\right)+3 r_\text{reg}\left(l^2\right)+2 l^4 r_\text{reg}''\left(l^2\right) \right) \\
    &\quad \quad -  \frac{1}{1 + r(l^2)} \left(11 l^2 r''\left(l^2\right)+10 r'\left(l^2\right)+2 l^4 r^{(3)}\left(l^2\right)\right) \bigg], \\
    I_{01}(r) &=  \frac{2 \pi}{3} \int_{l} \left(\frac{1}{1+r(l^2)} - \frac{r(l^2)}{l^2(1+r(l^2))^2} \right) \left(7 l^2 r_\text{reg}'\left(l^2\right)+3 r_\text{reg}\left(l^2\right)+2 l^4 r_\text{reg}''\left(l^2\right) \right), \\
    I_{12}(r) &= \frac{1}{\pi} \int_0^\infty dl \frac{-2l^2r'(l^2)}{(1+r(l^2))^2}, \\
    I_{02}(r) &= \frac{1}{\pi} \int_0^\infty dl \frac{r(l^2)}{(1+r(l^2))^2}, \\
    I_{15}(r) &= \int_{l_1 l_2 l_3} \frac{\dot{R}(l_{12}^2) - \dot{R}(l_{23}^2)}{P(l_1^2) P(l_2^2) P(l_3^2) P(l_{123}^2) P(l_{13}^2)} \left( l_1^2(l_2 \cdot l_3) - (l_1 \cdot l_3)(l_1 \cdot l_2) \right), \\
\label{eq:threshold05}
    I_{05}(r) &= \int_{l_1 l_2 l_3} \frac{R(l_{12}^2) - R(l_{23}^2)}{P(l_1^2) P(l_2^2) P(l_3^2) P(l_{123}^2) P(l_{13}^2)} \left( l_1^2(l_2 \cdot l_3) - (l_1 \cdot l_3)(l_1 \cdot l_2) \right) .
\end{align}
The notation $C_{i,j}$ makes clear the diagram topology giving rise to the contribution, with $i$ counting the number of insertions of $-p^4 r'(p^2/\kappa^2)$ (i.e., when the scale derivative in $\dot{R}$ hits the shape function) and $j$ counting the order of vertex times number of internal lines. In the main text, we have chosen to use a more digestible but less informative notation $C_{1,4} = a_1$, $C_{1,20} = a_2$, $C_{0,4} = b_1$, $C_{0,20} = b_2$. In the definition of the threshold integrals above, the function $r_\text{reg}(y)$ denotes the regular part of $r(y)$, defined as $r(y) = r_\text{reg}(y) + \frac{1}{y} \lim_{z\to 0} z r(z)$. Note that within our truncation, the WZW level $k$ stays quantised and is hence not (artificially) running, i.e., $\beta_k = 0$. In order to arrive at the above beta function, we have assumed that the shape function $r(y)$ is chosen with respect to the constraint
\begin{align}
    -y^4 r'(y) = C(r) + O(y^2),
    \label{eq:killnst}
\end{align}
which kills many of the non-standard contributions. Some of them would in fact be UV-divergent for arbitrary shape functions, and the presence of such contributions is not excluded \emph{a priori}, since the quartic regulator insertion does not regulate all momenta flowing through the insertion vertex independently, but only the combinations appearing in the Mandelstam variables, cf. Eqs.~\eqref{eq:quarticregvertex1}--\eqref{eq:quarticregvertex-1}. It is an open question whether all UV divergences can be removed merely by judicious choice of shape functions or by adding higher-order terms in the regulator insertion. Let us, however, hasten to add that it is in fact not strictly necessary to make the flow equation UV-finite in this way. Rather, a separate UV regularisation can be implemented to make the flow equations finite, the general framework for which has been worked out in \cite{Braun:2022mgx}.

\subsection{Regulator Scheme and Optimisation}

\begin{figure}[tb]
    \centering
    \includegraphics[scale=0.75]{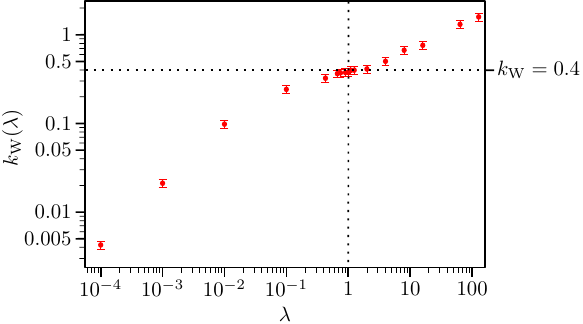}
    \caption{The critical WZW level $k_{\text{W}}$ for walking behaviour as function of the shape-function parameter $\lambda$, see Eq.~\eqref{eq:regsfam} for $A = A_\text{opt} = 0.65$. Since the level is formally a flow-invariant quantity, the point in regulator space where it is least sensitive to choice of regulator determines the optimal choice of regulator. This happens to be $\lambda \approx 1$.}
    \label{fig:kclambda}
\end{figure}

As usual when making an ansatz for the effective action, it is advisable to choose an optimised regulator which minimises the regulator-dependence of universal quantities. To this end, we consider the family of shape functions
\begin{align}
    r(y;\lambda, A) = \frac{\lambda}{y}e^{-A(y^2-1)}
    \label{eq:regsfam}
\end{align}
which fulfils the constraint in Eq.~\eqref{eq:killnst} for all $\lambda > 0$ and $A>0$. 
We first consider fixed $\lambda = 1$.
The family of shape functions $r(y;1,A)$ satisfies the normalisation condition $P_\kappa(k^2) = 2k^2$ for the regularised propagator.
An optimal value for the parameter $A$, yielding the largest radius of convergence for the derivative expansion, can then be obtained by maximising the gap $\inf_{p^2 > 0} P_\kappa(p^2)$~\cite{Litim:2001up}.
This way, we find $A_\text{opt} \approx 0.65$.
For this value of $A$, the family of shape functions $r(y; \lambda, A_\text{opt})$ as function of $\lambda$ satisfies another optimisation criterion, the principle of minimal sensitivity (PMS). As shown in Fig.~\ref{fig:kclambda}, the critical level for walking behaviour $k_{\text{W}}$ grows monotonically with $\lambda$, but features an inflection point located at $\lambda_\text{opt} \approx 1$.
At this value of $\lambda$, we therefore expect our predictions for universal quantities, such as $k_{\text{W}}$,  $k_{\text{NW}}$, and operator scaling dimensions, to be least sensitive to the choice of regulator, and as such yield the best-guess estimate at the given order of approximation.
In the numerical estimates given in the main text, we therefore show results using the optimised shape function $r(y; \lambda = 1, A = 0.65)$. The corresponding coefficients in the beta function, obtained from numerical integration of Eqs.~\eqref{eq:threshold11}--\eqref{eq:threshold05} using the \textsc{Cuhre} routine within the \textsc{Cuba} library~\cite{hahn:2016concurrent}, read
\begin{align}
    C_{0,4}^\text{opt} &\approx 0.22, &
    C_{1,4}^\text{opt} &\approx 0.90, \\
    C_{0,20}^\text{opt} &\approx 0.19, &
    C_{1,20}^\text{opt} &\approx 0.88.
\end{align}

\section{Acknowledgements}

We thank Zi Yang Meng, David Moser, Jan Pawlowski, Albert Agerholm Christensen and Shomrik Bhattacharya for useful discussions.
The work of L.J.\ is supported by the Deutsche Forschungsgemeinschaft (DFG) through SFB 1143 (A07, Project No.\ 247310070), the W\"urzburg-Dresden Cluster of Excellence \textit{ct.qmat} (EXC 2147, Project No.\ 390858490), and the Emmy Noether program (JA2306/4-1, Project No.\ 411750675). A.E. and S.R. are supported by a grant from VILLUM fonden (no.~29405). S.R. further acknowledges support from the DFG through the Walter Benjamin programme (RA3854/1-1, Project id No. 518075237).
M.M.S. acknowledges funding from the DFG within Project-ID 277146847, SFB 1238 (project C02), and the DFG Heisenberg programme (Project-ID 452976698).
B.H. and M.M.S. are supported by the Mercator Research Center Ruhr under Project No. Ko-2022-0012.

\bibliography{references}

\end{document}